\documentstyle[epsfig,aps]{revtex}
\begin{document}

\title{Output coupling and flow of a dilute Bose-Einstein condensate}

\author{B. Jackson, J. F. McCann, and C. S. Adams}
\address{Dept. of Physics, Rochester Building, University of Durham, South 
Road, Durham, DH1 3LE, UK.}

\date{\today}
\maketitle

\begin{abstract}

We solve the Gross-Pitaevskii equation to model the behaviour of a
weakly-interacting Bose condensate. Solutions are presented for the 
eigenstates in one- and two-dimensional harmonic traps. We include the effect 
of gravity and coupling to a second condensate to simulate an output coupler for
Bose condensed atoms, and find that the output pulse shape is in qualitative 
agreement with experiment. We also model flow under gravity through a 
constriction, demonstrating excitation and separation of eigenmodes in the 
condensate.

\end{abstract}

\section{Introduction}

The first experimental observations of Bose-Einstein condensation in
magnetically-trapped alkali atoms \cite{MHAN,KBDA,CCBR} excited
considerable interest in the properties of weakly-interacting Bose fluids.
Subsequent experiments have studied collective excitations \cite
{DSJI,MOME2} and sound propagation in the condensate \cite {MRAN,DMST},
leading to hopes of a deeper understanding of the nature of
superfluidity \cite {SSTR}. An important recent experiment has shown
macroscopic interference between two expanding condensates, providing
compelling evidence for coherence in the ground state of the trap \cite
{MRAN2}. Thus, the demonstration of an output coupler for Bose condensed
atoms \cite {MOME3} may be thought of as realizing a simple pulsed atom
laser, though controversy still surrounds this issue \cite {HMWI}.    

Theoretical work on weakly-interacting Bose condensates has focused on
solutions of the Gross-Pitaevskii or time-dependent nonlinear
Schr\"{o}dinger equation (NLSE) \cite{MEDW,PARU,MEDW2,MJHO}. This equation 
describes the evolution of a self-consistent wavefunction representing a 
dilute near-zero temperature condensate, and follows from mean-field theory 
where interactions between atoms are modelled by an $s$-wave scattering 
length \cite{NPPR}. The Gross-Pitaevskii equation has been shown to provide 
an accurate description of the shape \cite{MJHO} 
and dynamics \cite{DSJI,MOME2,MEDW3} of weakly interacting condensates, with
the principal effect of finite temperature arising as condensate depletion
\cite {RJDO}.                                    

The time-dependent NLSE may be solved numerically by a variety of methods 
\cite{TRTA}. Previous studies have used
semi-implicit Crank-Nicholson in one dimension \cite{PARU}, or
alternating-direction implicit-difference in two dimensions
\cite{MJHO}. In this work, we solve the NLSE using Split-Step Fourier 
\cite{TRTA,JJAV} and 
finite difference methods. A combination of these techniques can be used to 
model the dynamics of a condensate under a wide range of situations,  
for example, the creation of vortices by a moving object \cite{BJAC}. 

Here we consider two examples: simulations of the output coupling of Bose 
condensed atoms from a trap; and the flow of a condensate through a 
constriction. The output coupling of Bose condensed atoms, by driving an 
{\sc RF} transition to a untrapped state, has been demonstrated 
experimentally
by Mewes {\it et al.}\ \cite{MOME3}. Previous theoretical studies have
considered only 1D motion without gravity \cite{RJBA}, or have modelled the 
process by neglecting kinetic energy \cite{HSTE} or interaction terms \cite{WZHA} 
in the Gross-Pitaevskii equation. We solve coupled Gross-Pitaevskii 
equations in 2D, with the 
inclusion of gravity, to model the evolution of both the output pulse and the
remaining trapped condensate. The simulations reproduce the crescent-shaped 
atomic pulses observed experimentally \cite{MOME3,GTAU}, and we 
demonstrate that the recoil of the output pulse excites a harmonic 
oscillation of the parent condensate. Secondly, we consider techniques for 
condensate manipulation. To maintain the shape of a condensate pulse, it could 
be coupled into a waveguide such as a collimated far-off resonance laser beam. 
However, a more interesting situation arises in the flow through a constriction. 
For this case, we predict the excitation and spatial filtering of eigenmodes.
   
The paper is arranged as follows: Section 2 contrasts the Fourier and
finite-difference methods, and discusses 1D and 2D solutions
of the Gross-Pitaevskii equation for the lowest energy levels. Simulations of
the output coupling and flow through a constriction are presented in Sections
3 and 4 respectively.

\section{Numerical solution of the Gross-Pitaevskii equation}

The property of phase coherence inherent in Bose condensation implies that it 
may be represented by an order parameter, which is simply the macroscopic 
wavefunction of the condensate. At zero temperature this wavefunction,  
\( \Psi (\mbox{\boldmath $r$},t) \), is described by the Gross-Pitaevskii 
equation. Including a trap potential, $V(\mbox{\boldmath $r$},t)$, this has
the form \cite{PNOZ}:

\begin{equation}
i \hbar \frac{\partial}{\partial t} \Psi (\mbox{\boldmath $r$},t) = 
\left(- \frac{\hbar^{2}}{2m} \nabla^2 + V (\mbox{\boldmath $r$},t) + 
N U_0 |\Psi (\mbox{\boldmath $r$},t)|^2 \right) \Psi (\mbox{\boldmath $r$},t),
\label{eq:Gross-P}
\end{equation} 

where $m$ is the atomic mass, $N$ is the number of atoms, 
and $U_0$ describes the interaction between atoms in the condensate. For
$T \approx 0$ interactions are modelled by the $s$-wave scattering length, 
$a$, such that \(U_{0} = 4 \pi \hbar^{2} a/m \).

First, consider a 1D condensate, where the trapping potential is
described by \(V(x) = m\omega^{2} x^{2} /2 \). For convenience, equation 
(\ref{eq:Gross-P}) is scaled in terms of  harmonic oscillator units (h.o.u.) 
of length and time, where \(\xi = (\hbar/2m \omega)^{-1/2} x \) and 
\( \tau = \omega t \). With the change in length scale the normalisation of
the wavefunction is also modified, and thus 
\( \psi (\xi,\tau) = (\hbar / 2m\omega)^{1/4} \Psi (x,t) \). Taking the unit 
of energy to be $\hbar\omega$, equation (\ref{eq:Gross-P}) becomes 
\cite{PARU}:

\begin{equation}
i \frac{\partial}{\partial \tau} \psi (\xi,\tau) = \left(- \frac{\partial^2}
{\partial \xi^2} + \frac{1}{4} \xi^2 + C |\psi (\xi,\tau)|^2 \right) 
\psi (\xi,\tau),
\label{eq:NLSE}
\end{equation}

where \(C = 4\pi N a (2 \hbar / m \omega)^{1/2} \). In the next two 
subsections, we discuss the solution of this NLSE using (A) the Split-Step 
Fourier method, and (B) finite differences.

\subsection{Fourier method}

Consider a Schr\"{o}dinger equation of the form:

\begin{equation}
i\hbar \frac{\partial}{\partial t} \psi(\mbox{\boldmath $r$},t) = 
[T + V(\mbox{\boldmath $r$},t)] \psi(\mbox{\boldmath $r$},t) 
= H(\mbox{\boldmath $r$},t) \psi(\mbox{\boldmath $r$},t),
\end{equation}

with the solution, \( \psi (\mbox{\boldmath $r$},0) \), at $t=0$. We can 
propagate the solution through a short time interval to second-order 
acccuracy using the unitary evolution operator. Expansion of the 
time-evolution operator gives \cite{MDFE}:

\begin{equation}
 e^{-iHt/\hbar} \approx e^{-iV(0)t/2\hbar} e^{-iTt/\hbar} e^{-iV(0)t/2\hbar} +
 O (t^2).
\label{eq:half}
\end{equation}

For a slowly-varying $V(t)$, the order of accuracy of this `half-step'
expansion increases to $O (t^3)$. The operator, \(\exp(-iTt/\hbar) \) can be 
evaluated numerically by a variety of methods \cite{TRTA}. The Split-Step 
Fourier method relies on the use of
discrete Fourier transforms computed with a Fast Fourier Transform (FFT) 
algorithm, allowing high computational efficiency. Moreover, the Split-Step
Fourier method is readily extended to solution of the NLSE in an arbitrary 
number of dimensions by use of $N$-dimensional FFT routines \cite{WHPR}.

Solutions of (\ref{eq:NLSE}) may be evaluated numerically by beginning with 
an analytic ground or excited state solution in the absence of the nonlinear
term. The solution is then propagated through real time using the Split-Step 
Fourier method, while at each time step, the value of the nonlinear constant 
$C$ is increased adiabatically 
(`ramped') until reaching the desired value of $C$. In practice, this consists
of inserting a time-dependent pre-factor, 
\( P(\tau) = [1-\cos (\pi \tau / \tau_r)] /2 \), in front of the nonlinear
term \cite{MJHO}, where the ramp time $\tau_r$ is chosen to be long compared
to the oscillation time of the ground state of the trap. Alternatively, the
ground state can be readily determined through propagation in imaginary time
$\left(t \rightarrow -i\tilde{t} \right)$ \cite{JJAV}. However, we find that
imaginary time FFT is inappropriate for the excited
states, as these are exponentially suppressed with respect to lower energy modes.

One-dimensional wavefunctions for the first and second excited states are 
presented 
in Figs.\ 1 and 2, and show interesting behaviour as the repulsive 
interactions 
increase in strength. In both cases, the edges of the wavefunction spread out
in a similar fashion to the ground state \cite{PARU}. However, for the first 
excited state 
the width of the central node barely changes as $C$ 
increases; while in the second excited state, the two nodes of the symmetric
condensate asymptotically approach a fixed separation, 
\( \Delta \xi \sim 1 \), for large $C$. Later, we will demonstrate how such modes
could be excited by flow of a condensate through a constriction (Section 4).

\begin{figure}[htbp] 
\centering\epsfig{file=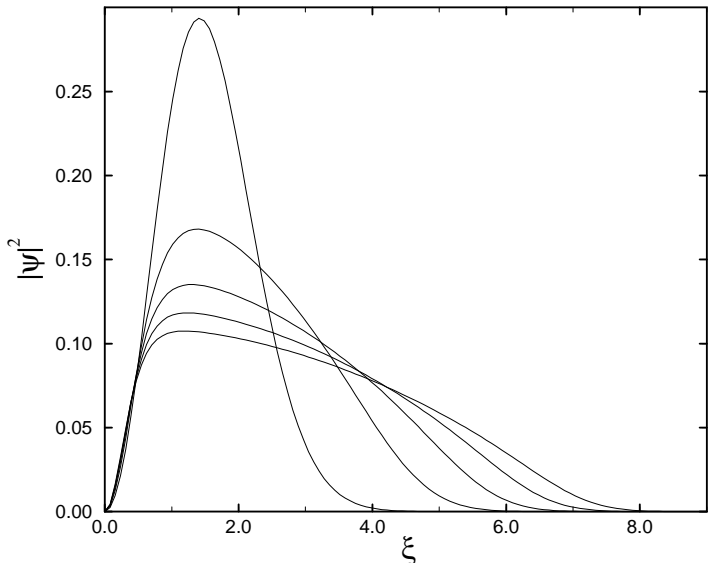, clip=, width=7.5cm, bbllx=100, bblly=40,
	 	  bburx= 550, bbury= 610, angle=270}
\caption{Condensate density $|\psi|^{2}$ plotted against position (in 
 h.o.u.) for the first excited state in 1D, with the nonlinear coefficient 
 $C=0,30,60,90,120$. As $C$  increases, the edges of the symmetric condensate
 spread out while the central node remains almost unchanged.}
\end{figure}

\begin{figure}[htbp]
\centering\epsfig{file=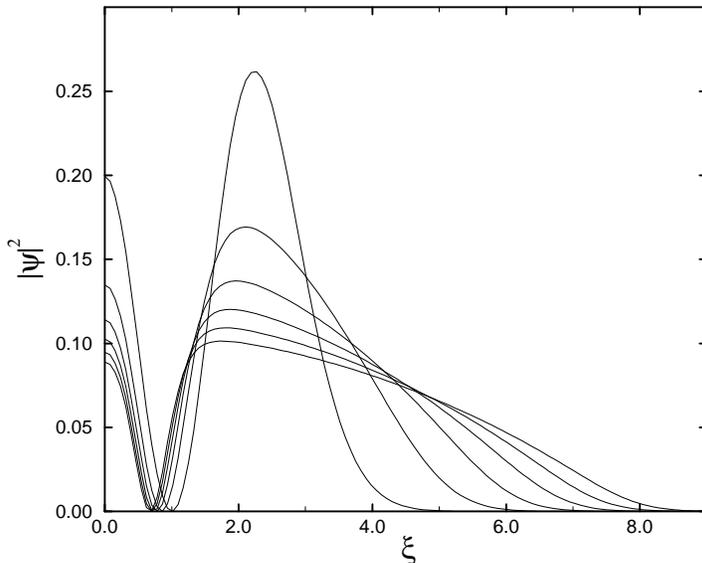, clip=, width=7.5cm, bbllx=100, 
 		  bblly=40, bburx= 550, bbury= 610, angle=270}
\caption{Condensate density $|\psi|^{2}$ plotted against position (in 
 h.o.u.) for the second excited state in 1D, for $C=0,30,60,90,120,150$. As 
 the interactions increase the nodes of the symmetric condensate converge, 
 and tend towards a fixed separation at high $C$.} 
\end{figure}

\subsection{Finite differences}

The procedure described above is difficult to implement for $C>150$, 
requiring long 
computation times---a problem compounded in 2D. An alternative approach is 
to 
evaluate the solutions of the NLSE and corresponding chemical potentials 
using finite differences. By replacing the differentials in 
(\ref{eq:NLSE}) by finite differences between all adjacent grid points, and 
including the condition that 
the wavefunction is normalised, we obtain a set of simultaneous equations for 
values of the wavefunction at each grid point. The simultaneous nonlinear
equations can then
be solved by iteration, e.g.\ using the Newton-Raphson method. Table I 
displays chemical potentials for the ground and first two excited states of 
the trap, in units 
of $\hbar \omega$. By neglecting the kinetic term in the time-independent 
NLSE---an approximation valid for large interaction strengths---one obtains a 
simple analytic solution for the ground state \cite{MEDW}, for which the 
chemical 
potential is $\mu_{1D}=(3C/8)^{2/3}$. This is commonly known as the
Thomas-Fermi approximation. At large $C$, the data for the ground state does 
indeed approach the Thomas-Fermi limit. In addition, the ground and excited 
energies converge so that they approach a fixed, equal separation at high $C$.

For a 2D isotropic trap, \( V(x,z)=m \omega^{2}(x^{2}+z^{2})/2 \), we can 
reduce
the problem to 1D by converting to cylindrical polar coordinates. Then the 
eigenvalue equation has the form:

\begin{equation}
\left(-\frac{\partial^{2}}{\partial \rho^{2}} -\frac{1}{\rho}\frac{\partial}
{\partial \rho}-\frac{1}{\rho^2}\frac{\partial^{2}}{\partial\varphi^{2}}
+\frac{1}{4}\rho^{2}-\mu+C|\psi(\rho,\varphi,\tau)|^{2} \right) 
\psi(\rho,\varphi,\tau)=0,
\end{equation}

where the ground state (i.e. no $\varphi$-dependence) is to be found. 
Ground-state chemical potentials for the 2D circular trap are presented in 
Table I. An important point to note is that the results approach the 
Thomas-Fermi limit, $ \mu_{2D} = (C/2\pi)^{1/2} $, more slowly than in 1D. 
The parabolic Thomas-Fermi
wavefunction fails to match the exact solution at the edges of the 
condensate. In 1D the wavefunctions mismatch at only two points, while in 2D
the mismatched region 
is a ring. Thus, as the dimensionality of the condensate increases, the 
Thomas-Fermi solution attains less accuracy, and one must be cautious when 
using this approximation in 3D.

This example also illustrates the ease with which the finite-difference 
method can find solutions of a NLSE containing large nonlinearities, when 
compared to the Split-Step Fourier method. However, our experience, in 
agreement with detailed studies \cite{TRTA}, indicates that the Fourier method
is more efficient when propagating the solution subject to a time-dependent 
potential. This is of particular significance when considering 2D 
simulations, where it is important that the computational times remain 
manageable. So, in the 2D simulations contained in this paper, we first find 
the ground-state solution of the NLSE for a trapped condensate using the
finite difference method, and then propagate the wavefunction under a 
time-dependent potential using the Split-Step Fourier method. 

\begin{table}\centering
\caption{Chemical potentials (in units of $\hbar \omega$) for the three 
 lowest energy states in 1D, and for the ground state of an isotropic 2D 
 condensate, calculated using finite differences. Ground state values are
 compared to the corresponding Thomas-Fermi
 values: $\mu_{1D}=(3C/8)^{2/3}$ and $\mu_{2D}=(C/2\pi)^{1/2}$.}
\begin{tabular}{ccccccc}                              	
	&\multicolumn{4}{c}{1D} & \multicolumn{2}{c}{2D} \\
	\cline{2-5} \cline{6-7}
$C$	&\multicolumn{2}{c}{Ground state} & \multicolumn{2}{c}{Excited states}
& \multicolumn{2}{c}{Ground state} \\ \cline{2-3} \cline{4-5} \cline{6-7}
	& Numerical solution & Thomas-Fermi & First & Second & Numerical 
solution & Thomas-Fermi \\	
0       & 0.500 & 0.000 & 1.500 & 2.500 & 1.000 & 0.000   \\
10      & 2.493 & 2.414 & 3.266 & 4.107 & 1.620 & 1.262   \\
20      & 3.886 & 3.832 & 4.630 & 5.427 & 2.064 & 1.784   \\
30      & 5.065 & 5.021 & 5.798 & 6.576 & 2.426 & 2.185   \\
50	& 7.091 & 7.058 & 7.816 & 8.576 & 3.020 & 2.821   \\
100	& 11.23	& 11.20 & 11.94 & 12.69 & 4.143 & 3.989	  \\	
200	& 17.80 & 17.78 & 18.51 & 19.24 & 5.760 & 5.642	  \\
500	& 32.77 & 32.76 & 33.48 & 34.20 & 9.003 & 8.921   \\ 
\end{tabular}
\end{table}

\section{The Output Coupler}

\subsection{The coupled Gross-Pitaevskii equations}
 
The previous section discussed how a combination of Fourier and finite 
difference methods are particularly efficient at finding and propagating 
solutions of the 
Gross-Pitaevskii equation. The same technique can be applied to mixtures of 
condensates. In particular, we solve coupled equations to describe the 
experimental realization of an output coupler for Bose condensed atoms, 
where a sequence of `daughter' condensate pulses is produced by excitation of a 
`parent' condensate \cite{MOME3}. For $^{23} {\rm Na}$ 
(or $^{87} {\rm Rb}$ \cite{MHAN}) 
condensates, the $F=1$ ground state is composed of the trapped state $M_F=-1$,
the magnetically insensitive level $M_F=0$, and the anti-trapped state 
$M_F=1$. Condensed atoms in each state may be represented by a Bose
wavefunction $\Psi_i (\mbox{\boldmath $r$},t)$, where $i\in\{-1,0,+1\}$. The
Hamiltonian of the system can then be constructed, to include gravitational
and magnetic trapping potentials, hyperfine interactions, and the {\sc RF}
coupling term  $ - \mbox{\boldmath $\mu$} \cdot \mbox{\boldmath $B$}
(\mbox{\boldmath $r$},t)$. Our approach involves applying the variational 
principle to obtain the following equations describing the dynamics of the
parent ($\Psi_{-1}$) and daughter condensates ($\Psi_0,\Psi_1$): 

\begin{equation}
 i \hbar {\partial \over \partial t} \Psi_k (\mbox{\boldmath $r$},t ) =
  \left[ -{\hbar^2 \over 2 m} \nabla^2 +U_k (\mbox{\boldmath $r$}) 
 + N \sum_{j} \langle \Psi_j \vert V \vert \Psi_j \rangle
 \right]  \Psi_k (\mbox{\boldmath $r$},t )
 + \hbar \sum_{j \neq k} \Omega_{kj} \Psi_j (\mbox{\boldmath $r$},t ),
\end{equation}
where $ U_k (\mbox{\boldmath $r$}) $ represents the gravitational and trap
potentials in the substate $k$, and $\Omega_{kj}$ is the Rabi frequency
describing coupling between sublevels. Atom-atom scattering appears as the 
matrix element $\langle \Psi_j \vert V \vert \Psi_j \rangle$. Assuming all
scattering lengths to be equal, $V$ may be replaced by the pseudopotential 
\( U_0 \delta (\mbox{\boldmath $r_i$} - \mbox{\boldmath $r_j$}) \).  

In the experiment, the {\sc RF} pulse was of short duration and small area 
\cite{MOME3}, leading to the following simplifications. First, due to the small 
area ($\vert \int \Omega_{+1,0} dt \vert \ll 1 $), coupling to the $M_{F}=-1$ 
state is small and may be safely ignored, and also back-coupling, 
i.e.\ Rabi oscillations, are negligible.  An example of strong coupling,
where Rabi oscillations are important, has been considered by Ballagh 
{\it et al.}\ \cite{RJBA} in a 1D weightless model.
Secondly, the short pulse duration, $T$ results in broadband excitation, so that
we may assume resonant coupling across the trap and neglect the spatial dependence
of the coupling term, $\Omega(\mbox{\boldmath $r$},t')$. In addition, we may 
assume that the parent condensate is `frozen' during the interaction 
($\omega T \ll 1$), in which case the daughter condensate is created suddenly as
a shadow of the parent, i.e.:
\begin{equation} 
 |\Psi_{0} (\mbox{\boldmath $r$},T)|^2 \approx \left |-i \int_{0}^{T} dt'
 {\Omega}(\mbox{\boldmath $r$},t')  
 \Psi_{-1} (\mbox{\boldmath $r$},t') \right|^2
 \approx f |\Psi_{-1} (\mbox{\boldmath $r$},0)|^2,
\label{eq:fracout_1}
\end{equation}
and,
\begin{equation}
 |\Psi_{-1}(\mbox{\boldmath $r$},T)|^2 \approx 
 (1-f)|\Psi_{-1}(\mbox{\boldmath $r$,0)}|^2,
\label{eq:fracout_2}
\end{equation} 
where $f$, the fraction of atoms coupled out of the trap, is equal to the
square of the pulse area. 

Following these simplifications, we obtain a pair of coupled time-dependent
equations in 2D:
\begin{equation} 
 i \partial_\tau \psi_{-1} = \left (- \nabla^2 + \textstyle{1 \over 4}
 (\xi^2+\eta^2) + C |\psi_{-1}|^2 + C |\psi_0|^2 \right) \psi_{-1},
\label{eq:couple_1}  
\end{equation}
\begin{equation}
 i \partial_\tau \psi_0 = \left (- \nabla^2 + G \eta + 
 C |\psi_0|^2 + C |\psi_{-1}|^2 \right) \psi_0,
\label{eq:couple_2}
\end{equation}
where $\xi=(\hbar/2m\omega)^{-1/2} x$, $\eta=(\hbar/2m\omega)^{-1/2} z$,
and $\psi_i(\xi,\eta,\tau)=(\hbar/2m\omega)^{-1/2}\Psi_i(x,z,t)$. Thus, 
$C= 8 \pi Na$, where $N$ is the linear density of atoms in the $y$-direction, and 
gravity is parametrized by $G=(m/2\hbar \omega^{3})^{1/2}g$. Note that gravity
is absent from Equation (\ref{eq:couple_1}), as this term only gives rise to
displacement of the equilibrium position of the trapped condensate. With 
initial conditions 
given by Equations  (\ref{eq:fracout_1}) and (\ref{eq:fracout_2}), equations
(\ref{eq:couple_1}) and (\ref{eq:couple_2}) are solved to model the
dynamical behaviour of the parent and daughter condensates for 
$\tau \geq \omega T$. The results of these simulations are now discussed.

\subsection{Results}

To simulate the MIT experiment we consider the parameters $C=200$ and $G=3.0$, 
corresponding to a trap frequency of 
\( \omega \simeq 2\pi \times 200\) Hz and $N \approx 2 \times 10^4$ 
$^{23} {\rm Na}$ atoms per h.o.u.\ respectively. Density plots of the output
pulse (for $f=0.2$) are presented in Fig. 3. After creation, the output pulse
is subject to gravity, a repulsive internal force, and a strong repulsion
from the residual parent condensate. 
We see in Fig. 3 that even at early times, a combination of these effects 
dramatically changes the shape of the output. The curvature of the daughter
condensate is found to be less pronounced for weaker interaction strengths.
At later times the output pulse expands freely, resulting in a
`crescent-shaped' density profile, 
characteristic of the MIT experiment \cite{MOME3,GTAU}. This suggests that 
the zero-temperature approximation, inherent in the coupled Gross-Pitaevskii 
equations, is appropriate for modelling the dynamics of the output.
The curvature of the output pulse was not observed in previous theoretical studies
\cite{WZHA}, due to a different trap geometry and subsequent neglect of the 
interaction term in the axial direction.

\begin{figure}[htbp]
\hspace{1.5cm}
\epsfig{file=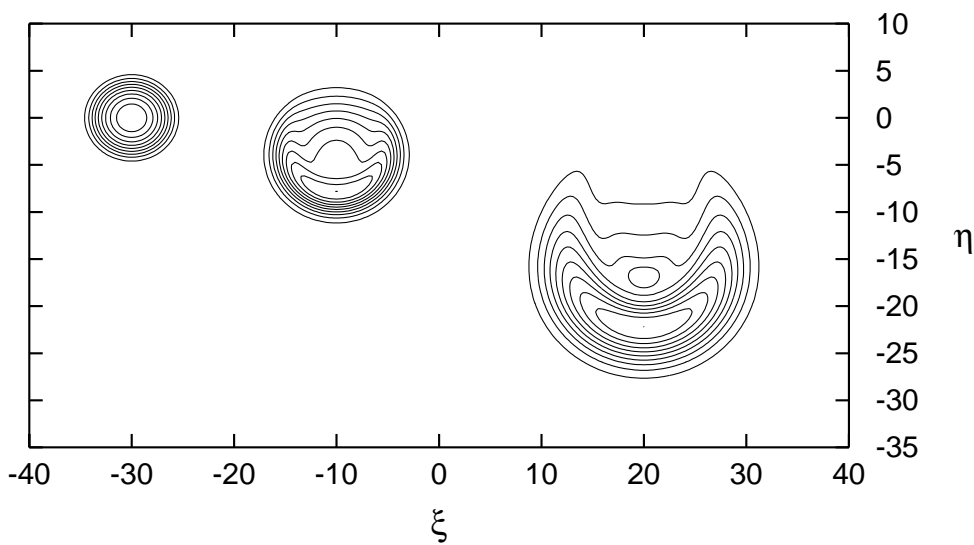, clip=,width=7.5cm, bbllx=495, bblly=205,
	bburx=230, bbury=665, angle = 90}
 \caption{Contour plot of $|\psi(\xi,\eta,\tau)|^{2}$ for an output 
 coupled Bose condensate falling under gravity, with $C=200$, $f=0.20$ and 
 $G=3.0$, at times $\tau=0$ (offset $\xi \rightarrow \xi-30$), $\tau=1.2$ 
 (offset $\xi\rightarrow \xi-10$), and $\tau=2.4$ (offset 
 $\xi\rightarrow\xi+20$). Nine equally-spaced contours are drawn for each 
 time-frame, up to peak densities of $|\psi(0,0,0)|^2 = 5.67 \times 10^{-3}$,
 $|\psi(0,-7.8125,1.2)|^2 = 2.36 \times 10^{-3}$,
 and $|\psi(0,-22.1875,2.4)|^2 = 1.25 \times 10^{-3}$. The repulsive
 force from the condensate remaining in the trap, combined with free 
 expansion under
 gravity, creates a `crescent-shaped' profile in the output.}
\end{figure}

The mutual repulsion between parent and daughter leads to a small recoil of 
the more massive, trapped condensate. The subsequent motion of the parent can
be inferred by studying its centre of mass, given by 
$\eta_c = \int\int\eta|\psi|^2 d\xi d\eta$. Simulations of a condensate in 
free-fall show that calculations of the centre of mass agree with the 
classical result, $\eta_c=-G \tau^2$, to five significant figures. The centre
of mass coordinate of the parent condensate along with its second time 
derivative (acceleration), are displayed
in Fig.\ 4. The repulsive force from the output pulse reaches a maximum at 
$\tau \sim 0.75$.
Subsequently, as the coupling term falls to zero due to the decreasing 
overlap, the parent undergoes harmonic motion at the trap frequency,
as expected.  Fig.\ 4 also illustrates that the recoil is less for smaller 
output fractions. Given that the output coupling for small $f$ does not 
seriously perturb the parent, further coherent pulses may be produced, and 
one would expect that the recoil on the parent would be negligible for slow, 
continuous output coupling.  

Given the agreement between these dynamical simulations and experiment, we surmise
that the same numerical methods can be employed to model the dynamics of the
condensate in other interesting situations. An example is discussed in the
following section. 

\begin{figure}[htbp]
\centering\epsfig{file=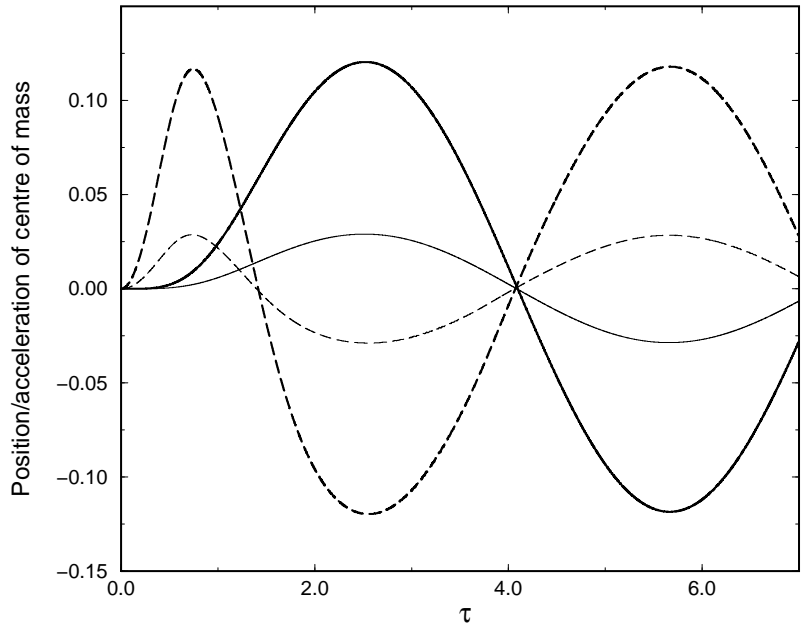, clip=,width=8.5cm, bbllx=90, 
bblly=30,
  bburx= 565, bbury= 635, angle=270}
 \caption{The vertical position (solid line) and acceleration 
 (dashed) of the centre of mass of the trapped parent condensate ($C=200$ and
 $G=3.0$) as a function of time. Repulsion from the output pulse induces a 
 recoil, which excites the parent into simple harmonic motion. Results are 
 presented for $f=0.20$ (thick lines) and $f=0.05$ (thin lines).} 
\end{figure}

\section{Flow through a constriction}

We now consider the flow of a weakly-interacting condensate through a
constriction. In the 2D simulations presented here, the condensate is released
from a circular trap and falls under gravity down a focused, far-off 
resonance
laser beam. The beam is approximated by a harmonic potential in which the 
spring constant varies with position along $z$. Any variation in 
potential along the $z$-axis (e.g.\ due to an increase in intensity near to 
the focus) is neglected. This situation may be realized using a blue-detuned 
doughnut
mode laser (see e.g.\ \cite{CSAD}). The model also assumes that there is no 
delay between release from the trap and switching on the laser beam. Under 
this model, for $\tau>0$, the evolution of the condensate is given by the 
equation:

\begin{equation}
i \partial_\tau \psi= \left(-\nabla^2 + {1 \over 4}
\left[\frac{1+(\eta_m^2/\eta_0^2)}{1+(\eta-\eta_m)^2/\eta_0^2}
\right]^2 \xi^2 + G \eta + C|\psi|^2 \right) \psi,
\end{equation}

so that the horizontal oscillation frequency, 
\( \omega=(\eta_0^2+\eta_m^2)/[\eta_0^2+(\eta-\eta_m)^2] \), 
is matched to the initial trap frequency at the `release' position, 
$\eta=0$, and increases by a factor of [\( 1+(\eta_{m}/\eta_{0})^{2} \)] at 
$\eta=\eta_{m}$.
The limit $\eta_0 \rightarrow \infty$ corresponds to a collimated laser beam.
In this case, the simulations show that the laser acts as a waveguide for 
output-coupled condensates.

An interesting situation arises when the laser is tightly-focused. Fig.\ 5 
displays results of the simulation for a small nonlinear coefficient 
($C=10$). The density plot at $\tau=2.0$ illustrates that three peaks are 
formed above the constriction. Scaling the 1D wavefunctions in Section 2 
illustrates that the peaks approximately match the positions of the maxima in 
Fig.\ 2, suggesting that the second excited state is populated and therefore 
parity is conserved. The probability that the condensate 
will be excited from the ground state is related to the rate of change
in the Hamiltonian compared to the excitation spectrum. For example, in 
the non-interacting limit ($C \rightarrow 0$), we can derive an 
approximate criterion for excitation to occur based on adiabaticity:
\( 4G^{1/2}|\eta_m|^{1/2}/\eta_0 \gg \Delta \mu \), 
where $\Delta \mu$ is the difference in energy between the ground and excited
states. Thus, a small constriction creates a sudden change in the Hamiltonian
and mode excitation occurs. Conversely, for wider constrictions, the 
Hamiltonian changes slowly and the fluid can adjust adiabatically.

\begin{figure}[htbp]
\hspace{4cm}
\epsfig{file=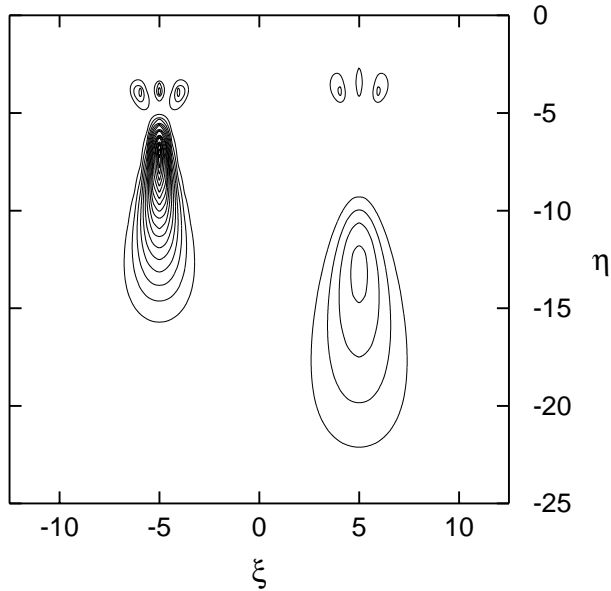, clip=,width=8cm, bbllx=475, bblly=165,
	bburx=160, bbury=490, angle=90}
 \caption{Flow of a weakly-interacting condensate ($C=10$, $G=3.0$) 
 through a constriction, where the oscillation frequency increases by a 
 factor of 11.5 between $\eta=0.0$ and $\eta_{m}=-6.0$. Contours 
 of the density are shown at $\tau=2.0$ (offset left) and $\tau=2.5$ (offset
 right).
 The excited `tail' of the condensate remains above the constriction while 
 the ground state falls under gravity, leading to spatial separation of the 
 eigenstates. The contours are drawn with an equal spacing of 
 $7 \times 10^{-3}$.}
\end{figure}

At subsequent times ($\tau=2.5$ in Fig. 5) the excited `tail' remains above 
the constriction, while the `bulk' of the condensate in the ground state 
continues to fall. To understand this separation of eigenmodes, consider 
the simple case of a non-interacting condensate, $C=0$. As the condensate 
falls into the constriction, the increase in oscillation frequency, 
$\delta \omega$, leads to a rise in the chemical
potential by $\delta \mu=\left(n+1/2 \right)\delta\omega$. This 
increase in energy is provided by a change in the gravitational potential,
$G\eta$. Equating these energies gives an expression for the 
equilibrium position, $\eta_e$:

\begin{equation}
\eta_e = \eta_m + \frac{\left(n+ \textstyle{1 \over 2} \right)}{2G}+
\sqrt{\frac{\left(n+\textstyle{1 \over 2} \right)^2}{4G^2}- 
\frac{\eta_m \left(n+\textstyle{1 \over 2} \right)}{G} - \eta_0^2},
\label{eq:separ1}
\end{equation}

So, for the $n$-th mode to be trapped, the following inequality must hold:

\begin{equation}
n \geq 2G \left(\eta_m + \sqrt{\eta_m^2 + \eta_0^2} \right) - \frac{1}{2}.
\label{eq:separ2}
\end{equation}

In the case of Fig.\ 5, equation (\ref{eq:separ1}) predicts 
$\eta_e \simeq -4.3$ for $n=2$, just below the actual position as
expected, because the small nonlinearity tends to push the mode upwards. 
In addition, the inequality (\ref{eq:separ2}) gives $n>1.2$, confirming
that only excited states are trapped above the constriction. The constriction
thus acts as a mode `filter', and could provide a convenient means to 
generate and study the decay of excitations. However, for a light-induced 
potential this would require tight-focusing: typically spot sizes of 1 
$\mu \rm{m}$ or less. 

For large $C$, the interaction term in the chemical potential dominates. 
Thus, in addition to the rise in energy due to the changing $\omega$, there 
is an increase 
in the density and therefore the interaction energy. As a result, the 
relative difference between eigenvalues is much lower than for $C=0$, and
mode separation 
tends not to occur. In Fig.\ 6, one can see that, for $C=200$, two maxima 
form above a wide constriction. Parity is still conserved, but the `tail' in 
this case consists mainly of a superposition of ground and second excited 
states. If narrower constrictions are used, many modes are 
excited, and the situation becomes increasingly complex.

\begin{figure}[htbp]
\hspace{4cm}
\epsfig{file=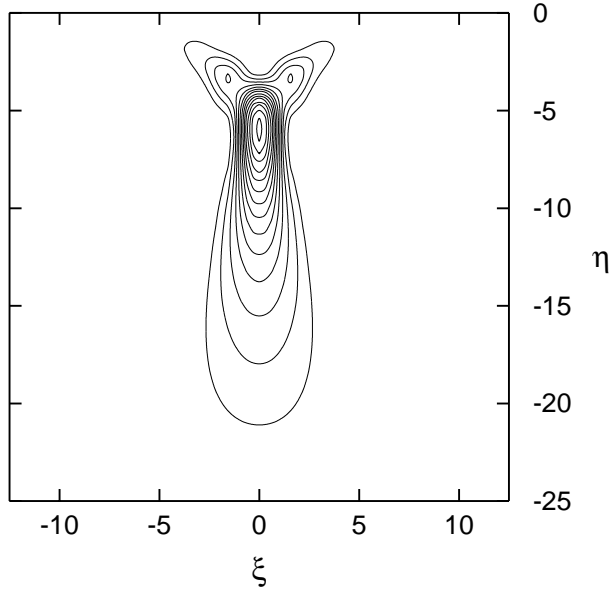, clip=,width=8cm, bbllx=475, bblly=165,
	bburx=160, bbury=490, angle=90}
 \caption{Flow of a condensate ($C=200$, $G=3.0$) through a 
 constriction at a time $\tau=2.0$, where the oscillation frequency increases
 by a factor of 2.5 between $\eta=0.0$ and $\eta=-6.0$. As for the 
 weakly-interacting case, the potential induces excitations in the 
 condensate. Contours are drawn with an equal spacing of 
 $3.03 \times 10^{-3}$.}  
\end{figure} 

\section{Summary}

In this paper, we present 1D and 2D solutions of the Gross-Pitaevskii equation. 
1D wavefunctions for the first and second
excited states are presented for various interaction strengths, and provide
useful information for the subsequent interpretation of our 2D simulations. 2D 
ground state solutions are also calculated for a circular geometry using a finite
difference method. We note that convergence towards the interaction dominated
Thomas-Fermi limit is slower in two than in one dimension.

The 2D wavefunctions were used as initial conditions in the following 2D
simulations. Firstly, an output coupler for magnetically-trapped
Bose condensed atoms was simulated by solution of a pair of coupled 
Gross-Pitaevskii
equations, to represent the output pulse and parent condensate. We demonstrate that
a combination of atomic interactions and gravity leads to curvature in the output,
in qualitative agreement with experiment. The simulations also demonstrate
that dipolar centre-of-mass modes are excited in the remaining trapped condensate.
The amplitude of the mode is particularly sensitive to the output fraction, and 
could provide a useful diagnostic in experiments. Secondly,
we presented simulations of the flow of a condensate dropped under gravity into a
tightly-focused laser beam. We demonstrate that a tight constriction may be used 
to excite and separate eigenmodes in a weakly-interacting
condensate. This could provide a useful technique to study the decay of excited 
condensate modes.

\acknowledgements
Financial support for this work was provided by the Nuffield Foundation and 
the EPSRC.

\end{document}